\documentclass[final,3p,times,twocolumn]{elsarticle}
 \biboptions{comma,sort&compress}
\usepackage{here}
 \usepackage{graphicx}
  \usepackage{epsfig}
\def\nin{\noindent}
\def\beq{\begin{equation}}
\def\eeq{\end{equation}}
\def\bea{\begin{eqnarray}}
\def\eea{\end{eqnarray}}

\journal{Nuc. Phys. (Proc. Suppl.)}

\begin{document}

\begin{frontmatter}



\title{On the magnetic catalysis and and inverse catalysis of phase transitions
in the linear sigma model}

 \author[label1,label3]{Alejandro Ayala,\fnref{Conacyt}}
  \address[label1]{Instituto de Ciencias Nucleares, Universidad Nacional Aut\'onoma de M\'exico,
  Apartado Postal 70-543, M\'exico Distrito Federal 04510, M\'exico}
  \address[label3]{Centre for Theoretical and Mathematical Physics and Department of Physics,
   University of Cape Town, Rondebosch 7700, South Africa} \ead{ayala@nucleares.unam.mx}

 \author[label2,label3]{M. Loewe,\corref{cor1}\fnref{Fondecyt1}}
  \address[label2]{Instituto de Física, Pontificia Universidad Cat\'olica de Chile,
  Casilla 306, Santiago 22, Chile} \ead{mloewe@fis.puc.cl}

\cortext[cor1]{Speaker}
\author[label4]{C. Villavicencio,\fnref{Fondecyt1}}
\address[label4]{Universidad Diego Portales, Casilla 298-V, Santiago, Chile} \ead{cvillavi@uc.cl}

\author[label2]{R. Zamora,\fnref{Doctoral}} \ead{rrzamora@puc.cl}

\fntext[Conacyt]{Supported by DGAPA-UNAM under grant
PAPIIIT-IN103811 and CONACyT under grant 128534}
\fntext[Fondecyt1]{Supported by Fondecyt under grant 1130056}
\fntext[Doctoral]{Supported by Fondecyt under grant 1130056 and
CONICYT under grant 21110295}

%\author{}

%\address{}

\begin{abstract}
\noindent We consider the evolution of critical temperature both for
the formation of a pion charged condensate as well as for the chiral
transition, from the perspective of the linear sigma model, in the
background of a magnetic field. We developed the discussion for the
pion condensate in one loop approximation for the effective
potential getting magnetic catalysis for high values of B, i.e a
raising of the critical temperature with the magnetic field. For the
analysis of the chiral restoration, we go beyond this approximation,
by taking one loop thermo-magnetic corrections to the couplings as
well as plasma screening effects for the boson's masses, expressed
through the resumation of ring diagrams. Here we found the opposite
behavior, i.e.  inverse magnetic catalysis, i.e. a decreasing of the
chiral critical temperature as function of the intensity of the
magnetic field, which seems to be in agreement with recent results
form the lattice community.

\end{abstract}


\end{frontmatter}

% \linenumbers

\section{Introduction}
%\label{}
\nin
%%%%%%%%%%%%
Recently \cite{LVZ} we studied the formation of a pion charged
condensate in the frame of  the linear sigma model. The main idea
was to consider the effective potential at the one loop level,
taking the isospin chemical potential near the effective pion mass,
varying then the intensity of the magnetic field in order to obtain
the critical temperature.

\noindent Defining

\begin{equation} \int_\beta dx \equiv \int_0^\beta
dx_4 \int d^3x,
\end{equation}

\noindent where $\beta=1/T$, being $T$ the temperature of the
system, using a Lagrangian without  fermionic sector but including
isospin chemical potential $\mu _{I}$  a magnetic field, through the
covariant derivative $ D_{\mu}=\partial_{\mu}+iqA_{\mu}$,and
expressed in terms of the charged pion fields $\pi_+$ and $\pi_-$,
the neutral pion field $\pi_0$ and the $\sigma$ field gives us the
following action

\begin{eqnarray}
S &=&  \int_\beta  dx \; \Big[
 (\partial_{4} - \mu_I)\pi_+(\partial_{4} + \mu_I)\pi_-
\nonumber \\ &&
\!\!+ (\partial_{i} - i eA_i)\pi_+  (\partial_{i} + i  eA_i)\pi_-  \nonumber \\
 &&\!\!+
\frac{1}{2} [(\partial \sigma)^2+ (\partial \pi_0)^2 +\mu_0^2 (\sigma^2+\pi_0^2+2\pi_+\pi_-)  ] \nonumber \\
                                                               &&+\frac{\lambda}{4}(\sigma^2+\vec{\pi}^2)^2-c\sigma \Big],
\label{lagrangian}
\end{eqnarray}

\smallskip
\noindent
 The term  $c\sigma$ corresponds to the explicit chiral symmetry breaking term,
 being $c=f_\pi m_\pi^2$ and where $f_\pi$
 is the pion decay constant.
 In the symmetric gauge, the external gauge field which produces a
 uniform magnetic field in the $z$ direction can be written as
\begin{equation}
\vec{A}=\frac{1}{2}\vec{B} \times \vec{r}= \frac{1}{2}B(-x_2,x_1,0).
\end{equation}

\noindent We may assume that the expectation value of the sigma
field $\bar{\sigma}$ has a non-vanishing value. This expectation
value, if the explicit term $c\sigma $ is absent is responsible for
the breaking of the chiral symmetry. Since the isospin symmetry is
also broken, due to the formation of the charged pion condensate, we
may expand both fields as quantum fluctuations around the classical
fields

\begin{equation} \sigma(x)=\bar{\sigma} + \widetilde{\sigma}(x), \hspace{3 mm}   \pi_\pm(x)= \frac{1}{\sqrt{2}} \phi_c(x)
 + \widetilde{\pi}_\pm(x).
\end{equation}

\noindent We proceed then to compute the effective potential, by
considering that the order parameter $\bar{v}$ , a spatial average
of the charged pion condensate, will be close to the normal phase,
i.e. $\bar{v} = 0$.

\begin{equation}
\bar v \equiv \left[\frac{1}{V}\int d^3 x \phi_c^2 \right]^{1/2},
\label{barv}
\end{equation}

\noindent where $V$ is the volume of the system. See \cite{LVZ} for
technical details. $\phi _{c}$ satisfies a non-relativistic
Schr\"{o}dinger equation

\begin{eqnarray}
[-\nabla^2+(eB)^2 (x_1^2+x_2^2)/4 + m_{\pi}^2-\mu_I^2] \phi_c = E^2
\phi_c.
\end{eqnarray}

\noindent which reminds us the two-dimensional harmonic oscillator
whose eigenvalues are given by

\begin{equation}
E_l^2(p_z)=p_z^2+m_{\pi}^2+(2l+1)eB - \mu_I^2.
\end{equation}

\begin{figure}[hbt]
\centerline{\includegraphics[width=7.cm]{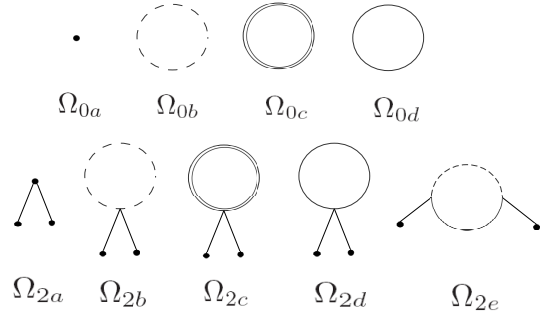}}
%{\epsfig{figure=mpsi2mc.eps,height=70mm}}
\caption{\scriptsize Relevant diagrams for the second derivative of
the effective potential with respect to the order parameter
$\bar{v}$ at $\bar{v}=0$.  The continuous line in the loop denotes
the sigma propagator, the dashed line is the charged pion
propagator, the double line represents  the neutral pion
 propagator,  and the
external lines represent $\phi_c$.}\label{fig0}
\end{figure}

\noindent From these considerations, and restricting ourselves to
the ground state, the classical field reads

\begin{equation}
\phi_c=\bar{v} \left( \frac{1-e^{-\Phi/2\Phi_0}}{\Phi/2\Phi_0}
\right)^{1/2} e^{-eB(x_1^2+x_2^2)/4}, \label{phi_c}
\end{equation}

\noindent where $\Phi\equiv BA$ is the magnetic flux, $A$ is the
area transverse to the external magnetic field and $\Phi_0 \equiv
\pi/q$ is the quantum magnetic flux. With this definition of the
order parameter $\bar{v}$ it turns out that the  tree-level
effective mass is independent of the magnetic flux. A different
prescription will produce a global flux dependent term. The relevant
Feynman diagrams for the computation of the effective potential are
shown in Fig.~\ref{fig0}

\nin

\begin{figure}[hbt]
\centerline{\includegraphics[width=8.cm]{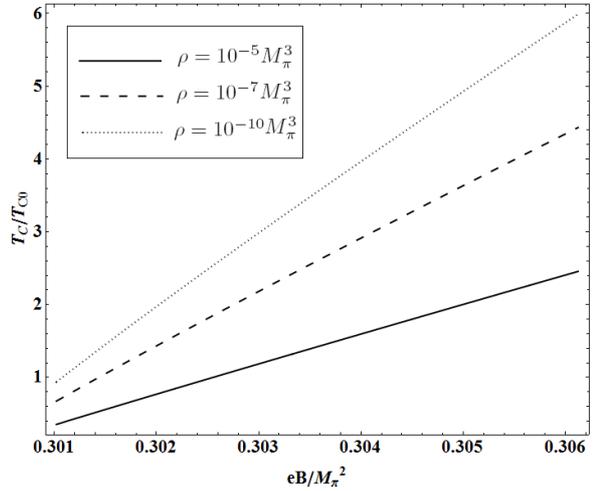}}
%{\epsfig{figure=mpsi2mc.eps,height=70mm}}
\caption{\scriptsize The critical temperature $T_{c}$ scaled with
the critical temperature in the absence of a magnetic field is shown
as a function of the magnetic field scaled with $M_{\pi
}^{2}$}\label{fig1}
\end{figure}

In this form we found, for high values of the magnetic field,
magnetic catalysis as it is shown in Fig.~\ref{fig1}. For lower
values of the magnetic field we found anti-catalysis.

%%%%%%%%%%%%%%%%%%%%%%%
\nin

%%%%%%%%%%%%
\section{Critical temperature for chiral restoration as function of the magnetic field}
%\label{}
\nin
%%%%%%%%%%%%

\noindent We also explored the chiral symmetry restoration, as
function of the temperature and the intensity of the magnetic field,
which is of course related to evolution of $\bar{\sigma}(T, B)$ but
going beyond the usual mean field approximation. In fact the
essential new points of our calculation are the introduction of
thermo-magnetic corrections to the vertices, both for  the bosonic
as well as for the fermionic sector, and the resumation of the self
energy corrections for the propagators of the bosonic field in the
evaluation of the effective potential. The charged pions and the
quark propagators were handled according to Schwinger's proper time
representation. This analysis has shown that an inverse magnetic
catalysis behavior, i.e the critical temperature decreases as
function of the magnetic field strength, emerges already for small
values of the magnetic field. In the next section we jus mention
some details. The reader should go to the original reference
\cite{sigma} for a complete description of the technical details.
Here we added the fermionic sector, neglecting density effects. Our
model is given then by

\begin{eqnarray}
   {\mathcal{L}}& = &\frac{1}{2}(\partial_\mu \sigma)^2  + \frac{1}{2}(D_\mu \vec{\pi})^2 + \frac{\mu^2}{2} (\sigma^2 +
   \vec{\pi}^2) \nonumber\\
   & - & \frac{\lambda}{4} (\sigma^2 + \vec{\pi}^2)^2
   + i \bar{\psi} \gamma^\mu D_\mu \psi -g\bar{\psi} (\sigma
   \nonumber\\
    & + & i \gamma_5 \vec{\tau} \cdot \vec{\pi} )\psi .
\label{lagrangian}
\end{eqnarray}

\begin{figure}[hbt]
\centerline{\includegraphics[width=6.cm]{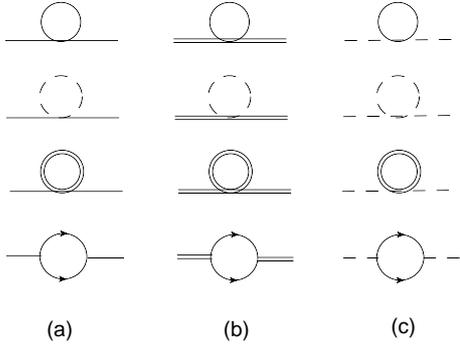}}
%{\epsfig{figure=mpsi2mc.eps,height=70mm}}
\caption{\scriptsize One-loop Feynman diagrams contribution to the
boson's self energies. The dashed line denotes the charged pion, the
continuous line is the sigma and the double line represents the
neutral pion and the continuous line with arrows represents the
quarks}\label{fig2}
\end{figure}

\noindent $\psi $ is an $SU(2)$ isospin doublet. We have computed
the effective potential going beyond the strictly one-loop
approximation by considering self energy leading high temperature
corrections to the boson propagators and one loop thermo-magnetic
corrections to the bosonic and fermionic couplings as well. The self
energy corrections were inserted later, according to the spirit of
the ring resumation, as plasma screening effects for the boson's
mass squared \cite{ayala et al}. The same philosophy was adopted
previously in a discussion of the abelian Higgs model \cite{abelian}
were were we also found magnetic anticatalysis.  In
Fig.~\ref{fig2}we show the relevant diagrams for the one loop boson
self energies. Only the charged pions feel the external magnetic
field, being then handled as Schwinger propagators.

\smallskip
\nin

\noindent The leading order at high temperature from these diagrams
is given by
\begin{eqnarray}
\Pi = \lambda \frac{T^{2}}{2} + N_{f} g^{2} \frac{T^{2}}{6},
\end{eqnarray}

\noindent where $N_{f}$ is the number of quark species.

\begin{figure}[hbt]
\centerline{\includegraphics[width=8.cm]{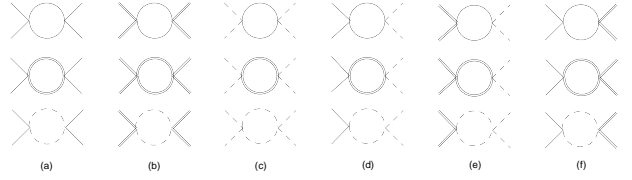}}
%{\epsfig{figure=mpsi2mc.eps,height=70mm}}
\caption{\scriptsize One-loop Feynman diagrams that contribute to
the thermo-magnetic corrections to the coupling $\lambda $. The
dashed line denotes the charged pion, the continuous line is the
sigma and the double line represents the neutral pion}\label{fig3}
\end{figure}

\begin{figure}[hbt]
\centerline{\includegraphics[width=5.cm]{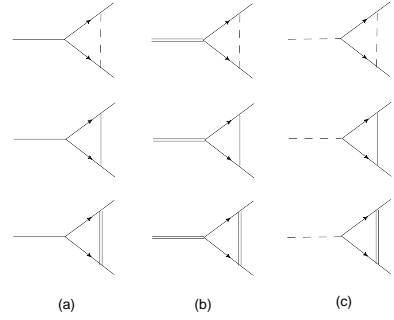}}
%{\epsfig{figure=mpsi2mc.eps,height=70mm}}
\caption{\scriptsize One-loop Feynman diagrams that contribute to
the thermo-magnetic corrections to the coupling g. The dashed line
denotes the charged pion, the continuous line is the sigma, the
double line represents the neutral pion and the continuous line with
arrows denote the quarks}\label{fig4}
\end{figure}

\noindent In Fig.~\ref{fig3} we show the relevant one loop diagrams
for the thermo-magnetic corrections to the coupling $\lambda $. In
Fig.~\ref{fig4} the relevant diagrams for the thermo-magnetic
corrections to the coupling g are shown.

\smallskip
\nin

\nin

\noindent The calculation for the coupling corrections were carried
on in the weak field limit approximation where well known
expressions for the bosonic \cite{mexicans} and the fermionic
propagators \cite{koreans} were used.

We refer the reader to \cite{sigma} for the explicit expression of
the effective coupling $\lambda _{eff} $ that emerges form this
analysis. It diminishes as function of the external magnetic field.
Contrary to the effective bosonic coupling, the triangle corrections
to the fermionic coupling $g$ produce an effective coupling
$g_{eff}$ that increases in a very mild way with the external field.

\begin{figure}[hbt]
\centerline{\includegraphics[width=8.cm]{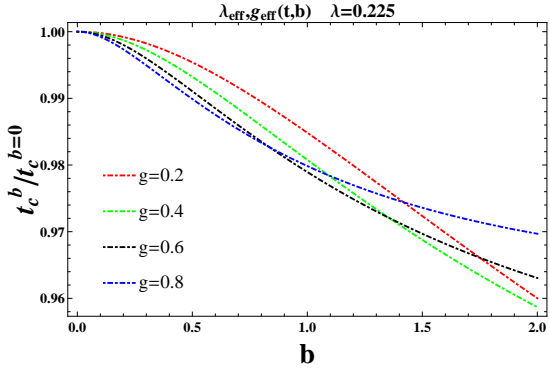}}
%{\epsfig{figure=mpsi2mc.eps,height=70mm}}
\caption{\scriptsize Color on-line Effect of the full thermomagnetic
dependence of couplings on the critical temperature for a fixed
value of the tree-level $\lambda = 0.225$ and different values of
the tree-level g as function of $b= qB/\mu ^{2}.$ In all cases the
critical temperature is a decreasing function of b}\label{fig5}
\end{figure}

\noindent The philosophy behind our calculation was to find the
effect of the magnetic field on the critical temperature where the
curvature of the effective potential vanishes. In Fig.~\ref{fig5} we
show the critical temperature behavior, normalized by the critical
temperature in the absence of an external field, including the full
thermo-magnetic dependence of the couplings. Here we have set the
tree level coupling $\lambda $ to fixed value and vary the tree
level coupling g.

\nin

\noindent A similar behavior was found in the complementary case
where we set the tree-level coupling g to a fixed value and vary the
tree-level coupling $\lambda $. These results seem to be in
agreement with recent lattice QCD results
\cite{Fodor,Bali:2012zg,Bali2} which indicate that the transition
temperature for chiral restoration with $2+1$ quark flavors, as
measured from the behavior of the chiral condensate and
susceptibility as well as from other thermodynamic observables,
significantly decreases with increasing magnetic field. It is
interesting to mention in this context a different approach.
Recently, Refs.~\cite{Farias, Ferreira1} have postulated
     an ad hoc magnetic field and temperature dependent running coupling, inspired by the QCD running of the coupling with energy,
     in the Nambu-Jona-Lasinio model, which makes the critical temperature decrease with increasing magnetic
     field.

%%%%%%%%%%%%%%%%
\section{Conclusions}

We have discussed the magnetic evolution of two different phase
transitions that may occur in hadronic physics: the formation of a
charged pion condensate triggered by the presence of a non-vanishing
isospin chemical potential, which is usually refer as the pion
superfluid phase and the restoration of chiral symmetry breaking.
The analysis was done in the frame of the linear sigma model. Our
results, from a strict one-loop analysis indicate magnetic catalysis
for the formation of the pion condensate for high values of the
magnetic field. The discussion of chiral symmetry breaking was much
more involved, since we incorporate in the effective potential
thermo-magnetic corrections to the couplings as well as plasma
screening effects emerging from a resumation of one loop self energy
boson corrections. Here we found anticatalysis, i.e. a decreasing
behavior of the critical temperature as function of the external
magnetic field. This result seems to be in agreement with recent
results form the lattice QCD community.

\nin
%%%%%%%%%%%%%%%%
%In a forthcoming work, we shall study the $X(3872)$ first observed by \cite{BELLE}.
%%%%%%%%%%%%%%%%%%%%%%%%%%%
\section*{Acknowledgements}
\noindent This work was supported in part by  DGAP-UNAM (M\'exico)
under grant No. PAPIT-IN103811 and CONACyT-M\'exico under grant No.
128534, FONDECYT (Chile) under grants No.1130056 and No. 1120770 and
CONICT (Chile) under grant No. 21110295. \nin

%%%%%%%%%%%%%%%%
%M.N. would like to thank the LPTA-Montpellier for hospitality.


%% References


%\bibliographystyle{elsarticle-num}





% \end{thebibliography}

%%%%%%%%%%%%%%%%%%%%
%\vfill\eject

%%%%%%%%%%%%%%%

\end{document}